\documentclass{aa}
\usepackage{graphicx}
\usepackage{amsmath,mathrsfs}
\usepackage{natbib}
\bibpunct{(}{)}{;}{a}{}{,}

\usepackage{txfonts}
\newcommand{\da}{\partial}
\newcommand{\alf}{Alfv\'{e}n }
\begin{document}

\title{On the validity of nonlinear Alfv\'{e}n resonance in space plasmas}

\author{C.~T.~M. Clack \and I. Ballai \and M.~S. Ruderman}

\offprints{I. Ballai}

\institute{Solar Physics and Space Plasma Research Centre
($SP^2RC$), Department of Applied Mathematics, University of
Sheffield, Hicks Building, Hounsfield Road, Sheffield, S3 7RH,
U.K.\\
\email{[app06ctc;i.ballai;m.s.ruderman]@sheffield.ac.uk}}

\date{Received 08 October 2008; accepted 21 November 2008}

\abstract{}{In the approximation of linear dissipative
magnetohydrodynamics (MHD) it can be shown that driven MHD waves in
magnetic plasmas with high Reynolds number exhibit a near resonant
behaviour if the frequency of the wave becomes equal to the local
\alf (or slow) frequency of a magnetic surface. This near resonant
behaviour is confined to a thin region, known as the dissipative
layer, which embraces the resonant magnetic surface. Although driven
MHD waves have small dimensionless amplitude far away from the
resonant surface, this near-resonant behaviour in the dissipative
layer may cause a breakdown of linear theory. Our aim is to study
the nonlinear effects in Alfv\'en dissipative layer}{In the present
paper, the method of simplified matched asymptotic expansions
developed for nonlinear slow resonant waves is used to describe
nonlinear effects inside the \alf dissipative layer.} {The nonlinear
corrections to resonant waves in the \alf dissipative layer are
derived and it is proved that at the \alf resonance (with
isotropic/anisotropic dissipation) wave dynamics can be described by
the linear theory with great accuracy.}{}

\titlerunning{Nonlinear effects in Alfv\'{e}n dissipative layers}
\authorrunning{Clack et al.}

\maketitle
%
%________________________________________________________________

\section{Introduction}

Magnetic fields are ubiquitously present in solar and space
plasmas. For regions where plasma-beta (the ratio of the kinetic
and magnetic pressures) is less than one, magnetism controls the
dynamics, topology and thermal state of the plasma. The magnetic
field in the solar atmosphere is not dispersed, but it tends to
accumulate in thinner or thicker entities often approximated as
magnetic flux tubes. These magnetic flux tubes serve as an ideal
medium for guided wave propagation.

One particular aspect of the solar physics that has attracted much
attention since the 1940s is the very high temperature of the solar
corona compared with the much cooler lower regions of the solar
atmosphere requesting the existence of some mechanism(s) which keeps
the solar corona hot against the radiative cooling. One of the
possible theories proposed is the transfer of omnipresent waves'
energy into thermal energy by resonant absorption or resonant
coupling of waves \citep[see e.g.][]{poedts1990, sakurai1,
goossens1995b}.

Waves which were initially observed sporadically mainly in radio
wavelengths \citep[see e.g.,][]{kai73, asch1992} are now observed in
abundance in all wavelengths, especially in (extreme) ultraviolet
\citep[see e.g.,][]{deforest98, asch1999, nak99, robb01, king03,
erd08, mar08}. Since the plasma is non-ideal, waves can lose their
energy through transport processes, however, the time over which the
waves dissipate their energy is far too long. In order to have an
effective and localized energy conversion, the plasma must exhibit
transversal inhomogeneities relative to the direction of the ambient
magnetic field. It was recognized a long time ago that solar and
space plasmas are inhomogeneous, with physical properties varying
over length scales much smaller than the scales determined by the
gravitational stratification. Homogenous plasmas have a
spectrum of linear eigenmodes which can be divided into slow, fast
and Alfv\'{e}n subspectra. The slow and fast subspectra have
discrete eigenmodes whereas the Alfv\'{e}n subspectrum is infinitely
degenerated. When an inhomogeneity is introduced the three
subspectra are changed. The infinite degeneracy of the Alfv\'{e}n
point spectrum is lifted and replaced by the Alfv\'{e}n continuum
along with the possibility of discrete Alfv\'{e}n modes occurring,
the accumulation point of the slow magnetoacoustic eigenvalues is
spread out into the slow continuum and a number of discrete slow
modes may occur, and the fast magnetoacoustic point spectrum
accumulates at infinity \citep[see e.g.,][]{goedbloed1975,
goedbloed1984}.

According to the accepted wave theories, effective energy transfer
between an energy carrying wave and the plasma occurs if the
frequency of the wave matches one of the frequencies in the slow or
\alf continua, i.e. at the slow or \alf resonances. The \alf
resonance has been more frequently associated with heating of
coronal structures given the low-$\beta$ regime of the solar corona.
Nevertheless, slow resonance cannot be ruled out as an additional
source of energy transfer. From a mathematical point of view, a
resonance is equivalent to regular singular points in the equations
describing the dynamics of waves, but these singularities can be
removed by, e.g., dissipation. Recently resonant absorption has
acquired a new applicability when the observed damping of waves and
oscillations in coronal loops has been attributed to resonant
absorption. Hence, resonant absorption has become a fundamental
constituent block of one of the newest branches of solar physics,
called \textit{coronal seismology} \citep[see e.g.,][]{nak99,
rudermanconf2002, goossens2002, arregui2007, ballai08, goossens08,
terradas2008} when applied to corona and solar magneto-seismology
when applied to the entire coupled solar atmosphere \citep[see
e.g.,][]{erdelyi07, verth07a, verth07b}.

Given the complexity of the mathematical approach, most theories
describing resonant waves are limited to the linear regime.
Perturbations, in these theories, are considered to be just small
deviations from an equilibrium despite the highly nonlinear
character of MHD equations describing the dynamics of waves and the
complicated interaction between waves and plasmas. Initial numerical
investigations of resonant waves in a nonlinear limit \citep[see e.g.,][]{ofman1995}
unveiled that the account of
nonlinearity introduces new physical effects which cannot be
described in the linear framework.

The first attempts to describe the nonlinear resonant waves
analytically appeared after the papers by \citet{ruderman1997b,
ruderman3} which were followed by further analysis by, e.g.,
\citet{Ballai1998a, Ballai1999}; \citet{Ruderman2000};
\citet{Clack2008}; however, all these papers focused on the slow
resonant waves only. These studies revealed that nonlinearity does
affect the absorption of waves. In addition, the absorption of
wave momentum generates a mean shear flow which can influence the
stability of resonant systems.

The present paper is the first analytical study on the nonlinear
resonant \alf wave, where we obtain governing equations using
techniques made familiar from previous studies on nonlinear slow
resonant MHD waves. Before embarking on the actual derivation, let
us carry out a qualitative discussion. First of all we should point
out that in plasmas with high Reynolds numbers (as in the solar
corona) efficient dissipation only operates in a thin layer
embracing the resonant surface. This layer is called \emph{the
dissipation layer}. This restriction on the effect of dissipation
makes the problem more tractable from a mathematical point of view,
as outside the dissipative layer the dynamics of waves is described
by the ideal MHD. Dissipation is a key ingredient of the problem of
resonance. As it was mentioned earlier dissipation removes
singularities in mathematical solutions. From a physical point of
view dissipation is important as it is the mechanism which relaxes
the accumulation of energy at the resonant surface and eventually
contributes to the global process of heating.

It is important to stress that the choice of dissipation has to be
related to the very physics which is described as different waves
are sensitive to different dissipative mechanisms. Due to the
dominant role of the magnetic field in the solar corona, transport
processes are highly anisotropic. Possible dissipation mechanisms
acting in coronal structures can be described within the framework
of Braginskii's theory \citep{braginskii} as it was shown in
applications by, e.g. \citet{erdelyi1995, ofman1995, mocanu2008}.
\alf waves are incompressible and transversal (in polarization),
therefore, it is sensible to adopt shear viscosity and magnetic
resistivity. Despite both transport processes being described by
rather small coefficients, the net effect of dissipation can be
increased considerably when the dissipative coefficients are
multiplied by large transversal gradients.

The paper is organized as follows. In the next section we introduce
the fundamental equations and discuss the main assumptions. In
Sect. III, we derive the governing equation for wave dynamics
inside the \alf dissipative layer. Section IV is devoted to
calculating the nonlinear corrections at \alf resonance. Finally, in Sect.
V we summarize and draw our conclusions, pointing out a few
applications and further studies to be carried out in the future.

\section{Fundamental equations and assumptions}

For describing mathematically the nonlinear resonant \alf waves we
use the visco-resistive MHD equations. In spite of the presence of
dissipation we use the adiabatic equation as an approximation of the
energy equation. Numerical studies by \citet{poedts1994}
in linear MHD have shown that dissipation due
to viscosity and finite electrical conductivity in the energy
equation does not alter significantly the behaviour of resonant
MHD waves in the driven problem.

When the product of the ion (electron) gyrofrequency,
$\omega_{i(e)}$, and the ion (electron) collision time,
$\tau_{i(e)}$, is much greater than one (as in the solar corona) the
viscosity and finite electrical conductivity become anisotropic and
viscosity is given by the Braginskii viscosity tensor (see Appendix
A). The components of the viscosity tensor that remove the
Alfv\'{e}n singularity are the shear components. The parallel and
perpendicular components of anisotropic finite electrical
conductivity only differ by a factor of $2$, therefore,
we will consider only one of them without loss of generality.

The dynamics of waves in our model is described by the
visco-resistive MHD equations
\begin{equation}
\frac{\da\bar{\rho}}{\da
t}+\nabla\cdot(\bar{\rho}\mathbf{v})=0,\label{eq:masscontinuity}
\end{equation}
\begin{equation}
\frac{\da\mathbf{v}}{\da
t}+(\mathbf{v}\cdot\nabla)\mathbf{v}=-\frac{1}{\bar{\rho}}\nabla\overline{P}+
\frac{1}{\mu_0\bar{\rho}}(\mathbf{B}\cdot\nabla)\mathbf{B}
+\frac{1}{\overline{\rho}}\nabla\cdot\mathbf{S},\label{eq:momentum}
\end{equation}
\begin{equation}
\frac{\da\mathbf{B}}{\da
t}=\nabla\times(\mathbf{v}\times\mathbf{B})+\overline{\lambda}\nabla^2\mathbf{B},\label{eq:induction}
\end{equation}
\begin{equation}\label{eq:adiabatic}
\frac{\da}{\da
t}\left(\frac{\overline{p}}{\overline{\rho}^{\gamma}}\right)+\mathbf{v}\cdot\nabla\left(\frac{\overline{p}}{\overline{\rho}^{\gamma}}\right)=0,
\end{equation}
\begin{equation}
\overline{P}=\bar{p}+\frac{\mathbf{B}^2}{2\mu_0},\mbox{
}\nabla\cdot\mathbf{B}=0.\label{eq:totalpressureandsolenoid}
\end{equation}
In Eqs.
(\ref{eq:masscontinuity})-(\ref{eq:totalpressureandsolenoid})
$\mathbf{v}$ and $\mathbf{B}$ are the velocity and magnetic
induction vectors, $\bar{p}$ the plasma pressure, $\bar{\rho}$ the
density, $\overline{\lambda}$ the coefficient of magnetic
diffusivity, $\gamma$ the adiabatic exponent, and $\mu_0$ the
magnetic permeability of free space. In addition,
$\nabla\cdot\mathbf{S}$ is the Braginskii viscosity (see Appendix A
for full details). Note that even though anisotropy has been
considered the Hall term has been neglected. We neglect the Hall
term from the induction equation, which can be of the order of
diffusion term in the solar corona, because the largest Hall terms
in the perpendicular direction relative to the ambient magnetic
field identically cancel. The components of the Hall term in the
normal and parallel directions relative to the ambient magnetic
field have no effect on the dynamics of \alf waves in dissipative
layers, hence these too are neglected. For full details on the Hall
term and the reasoning behind neglecting it, we refer to Appendix B.

We adopt Cartesian coordinates $x, y, z$ and limit our analysis to a
static background equilibrium ($\mathbf{v}_0=0$). We assume that all
equilibrium quantities depend on $x$ only. The equilibrium magnetic
field, $\mathbf{B}_0$, is unidirectional and lies in the $yz$-plane.
The equilibrium quantities must satisfy the condition of total
pressure balance,
\begin{equation}\label{eq:pressurebalance}
p_0+\frac{B_{0}^2}{2\mu_0}=\rm{constant}.
\end{equation}
For simplicity we assume that the perturbations of all quantities
are independent of $y$ ($\da/\da y =0$). We note that since the
magnetic field is not aligned with the $z$-axis, Alfv\'{e}n waves
still exist. In linear theory of driven waves all perturbed
quantities oscillate with the same frequency, $\omega$, which means
that they can be Fourier-analyzed and taken to be proportional to
$\exp(i[kz-\omega t])$. Solutions are sought in the form of
propagating waves. All perturbations in these solutions depend on
the combination $\theta=z-Vt$, rather than $z$ and $t$ separately,
with $V=\omega/k$. In order to match linear theory as closely as
possible we apply the same procedure as above. In the context of
resonant absorption the phase velocity, $V$, must match the
projection of the Alfv\'{e}n velocity, $v_A$, onto the $z$-axis when
$x=x_A$ where $x_A$ is the resonant position. To define the resonant
position mathematically it is convenient to introduce the angle,
$\alpha$, between the $z$-axis and the direction of the equilibrium
magnetic field, so that the components of the equilibrium magnetic
field are
\begin{equation}\label{eq:angles}
B_{0y}=B_0\sin\alpha,\phantom{X}B_{0z}=B_0\cos\alpha.
\end{equation}
The definition of the resonant position can now be written mathematically as
\begin{equation}\label{eq:resonantposition}
V=v_{A}\left(x_A\right)\cos\alpha,
\end{equation}
where $v_A$ is the \alf speed defined as
\begin{equation}\label{eq:alfvensound}
v_A=\frac{B_0}{\sqrt{\mu_0\rho_0}}.
\end{equation}
In addition we introduce the sound and cusp speeds as
\begin{equation}\label{eq:cuspspeed}
c_S=\left(\frac{\gamma p_0}{\rho_0}\right)^{1/2},\mbox{ }
c_T=\left(\frac{c_S^2v_A^2}{c_S^2+v_A^2}\right)^{1/2}.
\end{equation}
In what follows we can take $x_A=0$ without loss of generality. The
perturbations of the physical quantities are defined by
\begin{align}
&\bar{\rho}=\rho_0+\rho,\phantom{X} \bar{p}=p_0+p,\phantom{X}
\mathbf{B}=\mathbf{B}_0+\mathbf{b},\nonumber\\
&P=p+\frac{\mathbf{B}_0\cdot\mathbf{b}}{\mu_0}+\frac{\mathbf{b}^2}{2\mu_0},\label{eq:perturbations}
\end{align}
where $P$ is the perturbation of total pressure.

The dominant dynamics of resonant \alf waves, in linear MHD, resides
in the components of the perturbed magnetic field and velocity that
are perpendicular to the equilibrium magnetic field and to the
$x$-direction. This dominant behaviour is created by an $x^{-1}$
singularity in the spatial solution of these quantities at the \alf
resonance \citep{sakurai1, goossens1995a}; these variables are known
as \emph{large variables}. The $x$- component of velocity, the
components of magnetic field normal and parallel to the equilibrium
magnetic field, plasma pressure and density are also singular,
however, their singularity is proportional to $\ln|x|$. In addition,
the quantities $P$ and the components of $\mathbf{v}$ and
$\mathbf{b}$ that are parallel to the equilibrium magnetic field are
regular; all these variables are called \emph{small variables}.

To make the mathematical analysis more concise and the physics more
transparent we define the components of velocity and magnetic field
that are in the $yz$-plane and are either parallel or perpendicular to the equilibrium magnetic
field:
\begin{align}
&\left(
\begin{array}{cc}
v_{\parallel}\\
b_{\parallel}
\end{array}\right)=\left(
\begin{array}{cc}
v\phantom{X} w\\
b_y\phantom{X} b_z
\end{array}\right)
\left(\begin{array}{cc}
\sin\alpha\\
\cos\alpha
\end{array}\right),\nonumber\\
&\left(
\begin{array}{cc}
v_{\perp}\\
b_{\perp}
\end{array}\right)=\left(
\begin{array}{cc}
v\phantom{X} -w\\
b_y\phantom{X} -b_z
\end{array}\right)
\left(\begin{array}{cc}
\cos\alpha\\
\sin\alpha
\end{array}\right),
\end{align}
where $v$, $w$, $b_y$ and $b_z$ are the $y$- and $z$-components of
the velocity and perturbation of magnetic field, respectively.

Let us introduce the characteristic scale of inhomogeneity,
$l_{inh}$. The classical viscous Reynolds number, $R_e$, and the
magnetic Reynolds number, $R_m$, are defined as
\begin{equation}\label{eq:reynoldsnumbers}
R_e=\frac{\widetilde{\rho}_0Vl_{inh}}{\overline{\eta}},\phantom{x}R_m=\frac{Vl_{inh}}{\bar{\lambda}},
\end{equation}
where $\widetilde{\rho}_0$ is a characteristic value of $\rho_0$, and $\overline{\eta}=\overline{\eta}_1$ is the
shear viscosity coefficient (see Appendix A).
These two numbers determine the importance of viscosity and finite electrical conductivity. We introduce the total Reynolds number as
\begin{equation}\label{eq:totalreynolds}
\frac{1}{R}=\frac{1}{R_e}+\frac{1}{R_m}.
\end{equation}
The aim of this paper is to study the nonlinear behaviour of driven
\alf resonant waves in the dissipative layer. We are not interested
in MHD waves that have large amplitude everywhere and require a
nonlinear description in the whole space. We focus on waves that
have small dimensionless amplitude $\epsilon\ll 1$ far away from the
ideal \alf resonant point $x=0$.

In nonlinear theory, when studying resonant behaviour in the
dissipative layer we must scale the dissipative coefficients
\citep[see e.g.,][]{ruderman3, Ballai1998a, Clack2008}. The
general scaling to be applied is
\begin{equation}\label{eq:reycoefscales}
\overline{\eta}=R^{-1}\eta,\mbox{ }\overline{\lambda}=R^{-1}\lambda.
\end{equation}
Linear theory predicts that the characteristic thickness of the
dissipative layer, $l_{diss}$, is of the order of $l_{inh}R^{-1/3}$
and we assume that this is true in the nonlinear regime, too. Hence,
we must introduce a stretching transversal coordinate, $\xi$, in the
dissipative layer defined as
\begin{equation}\label{eq:transveralscale}
\xi=R^{1/3}x.
\end{equation}

We can rewrite Eqs.
(\ref{eq:masscontinuity})-(\ref{eq:totalpressureandsolenoid}) in
the scalar form as
\begin{equation}\label{eq:masscontinuity1}
V\frac{\da \rho}{\da\theta}-\frac{\da(\rho_0 u)}{\da
x}-\rho_0\frac{\da w}{\da\theta}=\frac{\da(\rho u)}{\da
x}+\frac{\da(\rho w)}{\da\theta},
\end{equation}
\begin{multline}\label{eq:momentumx1}
\rho_0 V\frac{\da u}{\da\theta}-\frac{\da P}{\da
x}+\frac{B_0 \cos\alpha}{\mu_0}\frac{\da b_x}{\da\theta}=
\bar{\rho}\left(u\frac{\da u}{\da x}+w\frac{\da
w}{\da\theta}\right)\\
-\rho V\frac{\da u}{\da\theta}-\frac{b_x}{\mu_0}\frac{\da b_x}{\da
x}-\frac{b_z}{\mu_0}\frac{\da b_x}{\da\theta}
-\overline{\eta}\frac{\da^2 u}{\da x^2},
\end{multline}
\begin{multline}\label{eq:momentumperp1}
\frac{\da}{\da\theta}\left(\rho_0 V
v_{\perp}+P\sin\alpha+\frac{B_0\cos\alpha}{\mu_0}b_{\perp}\right)
=\bar{\rho}\left(u\frac{\da v_{\perp}}{\da x}+w\frac{\da
v_{\perp}}{\da\theta}\right)\\
-\rho V\frac{\da
v_{\perp}}{\da\theta}
-\frac{b_x}{\mu_0}\frac{\da b_{\perp}}{\da
x}-\frac{b_z}{\mu_0}\frac{\da b_{\perp}}{\da\theta}
-\overline{\eta}\frac{\da^2 v_{\perp}}{\da x^2},
\end{multline}
\begin{multline}\label{eq:momentumpara1}
\frac{\da}{\da\theta}\left(\rho_0 V
v_{\parallel}-P\cos\alpha+\frac{B_0\cos\alpha}{\mu_0}b_{\parallel}\right)
=\bar{\rho}\left(u\frac{\da v_{\parallel}}{\da x}+w\frac{\da
v_{\parallel}}{\da\theta}\right)\\
-\frac{b_x}{\mu_0}\frac{dB_0}{dx}-\rho V\frac{\da
v_{\parallel}}{\da\theta}
-\frac{b_x}{\mu_0}\frac{\da b_{\parallel}}{\da
x}-\frac{b_z}{\mu_0}\frac{\da b_{\parallel}}{\da\theta}
-4\overline{\eta}\frac{\da^2 v_{\parallel}}{\da x^2},
\end{multline}
\begin{equation}\label{eq:inductx1}
Vb_x+B_0 u\cos\alpha=wb_x-ub_z +\overline{\lambda}\left(\frac{\da
b_x}{\da\theta}-\frac{\da b_z}{\da x}\right),
\end{equation}
\begin{multline}\label{eq:inductperp1}
\frac{\da}{\da\theta}\left(Vb_{\perp}+B_0
v_{\perp}\cos\alpha\right)=\frac{\da(ub_{\perp})}{\da
x}+\frac{\da(wb_{\perp})}{\da\theta}\\
-b_x\frac{\da v_\perp}{\da x}-b_z\frac{\da v_{\perp}}{\da\theta}
-\overline{\lambda}\nabla^2 b_{\perp},
\end{multline}
\begin{multline}\label{eq:inductpara1}
\frac{\da}{\da\theta}\left(Vb_{\parallel}+B_0
v_{\parallel}\cos\alpha\right)-\frac{\da(B_0 u)}{\da
x}-B_0\frac{\da w}{\da\theta}\\
=\frac{\da(ub_{\parallel})}{\da
x}+\frac{\da(wb_{\parallel})}{\da\theta}-b_x\frac{\da
v_{\parallel}}{\da x}-b_z\frac{\da v_{\parallel}}{\da\theta}
-\overline{\lambda}\nabla^2 b_{\parallel},
\end{multline}
\begin{multline}
V\left(\frac{\da p}{\da\theta}-c_S^2\frac{\da \rho}{\da\theta}\right)-
u\left(\frac{dp_0}{dx}-c_S^2\frac{d\rho_0}{dx}\right)\\
=\frac{1}{\rho_0}\left\{V\left(\gamma p\frac{\da \rho}{\da \theta}-\rho\frac{\da p}{\da\theta}\right)
-w\left[\gamma\overline{p}\frac{\da\rho}{\da\theta}-\overline{p}\frac{\da p}{\da\theta}\right]\right.\\
\left.+u\left[\rho\frac{dp_0}{dx}-\gamma p\frac{d\rho_0}{dx}+\overline{\rho}\frac{\da p}{\da x}
-\gamma\overline{p}\frac{\da \rho}{\da x}\right]\right\}
\end{multline}
\begin{equation}\label{eq:totalpressure1}
P=p+\frac{1}{2\mu_0}\left(b_x^2+b_{\perp}^2+b_{\parallel}^2+2B_0
b_{\parallel}\right),
\end{equation}
\begin{equation}\label{eq:solenoid1}
\frac{\da b_x}{\da x}+\frac{\da b_z}{\da\theta}=0.
\end{equation}
In the above equations $\nabla=(\da/\da x,0,\da/\da\theta)$ and
$w=v_{\parallel}\cos\alpha-v_{\perp}\sin\alpha$.

Equations (\ref{eq:masscontinuity1})-(\ref{eq:solenoid1}) will be used in the following sections
to derive the governing equation for the resonant \alf waves inside the dissipative layer and to
find the nonlinear corrections.

\section{The governing equation in the dissipative layer}

In order to derive the governing equation for wave motions in the
\alf dissipative layer we employ the method of matched asymptotic
expansions \citep{nayfeh1981, bender1999}. This method requires to
find the so-called \textit{outer} and \textit{inner} expansions
and then match them in the overlap regions. This nomenclature is
ideal for our situation. The outer expansion corresponds to the
solution outside the dissipative layer and the inner expansion
corresponds to the solution inside the dissipative layer. A
simplified version of the method of matched asymptotic expansions,
developed by \citet{Ballai1998a}, is adopted here.

The typical largest \emph{quadratic} nonlinear term in the system of
MHD equations is of the form $g\da g/\da z$ while the typical
dissipative term is of the form $\overline{\eta}\da^2 g/\da z^2$,
where $g$ is any `large' variable. Linear theory predicts that
`large' variables have an ideal singularity $x^{-1}$ in the vicinity
of $x=0$. This implies that the `large' variables have dimensionless
amplitudes in the dissipative layer of the order of $\epsilon
R^{1/3}$. It is now straightforward to estimate the ratio of a
typical quadratic nonlinear and dissipative term,
\begin{equation}\label{eq:ratioofnonlindiss}
\phi_q=\frac{g\da g/\da z}{\overline{\eta}\da^2 g/\da
z^2}=\mathscr{O}(\epsilon R^{2/3}),
\end{equation}
where the quantity $\phi_q$ can be considered as the \emph{quadratic
nonlinearity parameter}. If the condition $\epsilon R^{2/3}\ll1$ is
satisfied, linear theory is applicable. On the other hand, if
$\epsilon R^{2/3}\gtrsim 1$ then nonlinearity has to be taken into
account when studying resonant waves in dissipative layers. Using
the same scalings, \citet{ruderman3} showed that nonlinearity has to
be considered whenever slow resonant waves are studied in the solar
photosphere. For a typical dimensionless amplitude of
$\epsilon\sim10^{-2}$ linear theory can be applied if the total
Reynolds number is less than $10^3$. This value is much less than
the resistive and shear viscosity Reynolds number
($10^{10}-10^{12}$). This conclusion implies that in the solar
atmosphere resonant absorption should be a nonlinear phenomenon. In
order to describe the role of dissipation and nonlinearity equally
we assume that $\phi_q\sim1$.

Far away from the dissipative layer the amplitudes of perturbations
are small, so we use linear ideal MHD equations in order to describe
the wave motion. The full set of nonlinear dissipative MHD equations
are used for describing wave motion \textit{inside} the dissipative
layer where the amplitudes can be large. We, therefore, look for
solutions in the form of asymptotic expansions. The equilibrium
quantities change only slightly across the dissipative layer so it
is possible to approximate them by the first non-vanishing term in
their Taylor series expansion with respect to $x$. Similar to linear
theory, we assume that the expansions of equilibrium quantities are
valid in a region embracing the ideal resonant position, which is
assumed to be much wider than the dissipative layer. This implies
that there are two overlap regions, one to the left and one to the
right of the dissipative layer, where both the outer (the solution
to the linear ideal MHD equations) and inner (the solution to the
nonlinear dissipative MHD equations) solutions are valid. Hence,
both solutions must coincide in the overlap regions which provides
the matching conditions.

Before deriving the nonlinear governing equation we ought to make a
note. In linear theory, perturbations of physical quantities are
harmonic functions of $\theta$ and their mean values over a period
are zero. In nonlinear theory, however, the perturbations of
variables can have non-zero mean values as a result of nonlinear
interaction of different harmonics. Due to the absorption of wave
momentum, a mean shear flow is generated outside the dissipative
layer \citep{ofman1995}. This result is true for our analysis also,
however, due to the length of this study we prefer to deal with this
problem in a forthcoming paper.

We suppose that nonlinearity and dissipation are of the same order
so we have $\epsilon R^{2/3}=\mathscr{O}(1)$, i.e. $R\sim
\epsilon^{-3/2}$. We can, therefore, substitute $\epsilon^{-3/2}$
for $R$ in Eq. (\ref{eq:reycoefscales}) to rescale viscosity and
finite electrical resistivity as
\begin{equation}\label{eq:rescalecoeffquadratic}
\overline{\eta}=\epsilon^{3/2}\eta,\mbox{
}\overline{\lambda}=\epsilon^{3/2}\lambda.
\end{equation}
We do not rewrite the MHD equations as they are easily obtained from
Eqs. (\ref{eq:masscontinuity1})-(\ref{eq:solenoid1}) by substitution
of Eq. (\ref{eq:rescalecoeffquadratic}).

The first step in our description is the derivation of governing
equations outside the dissipative layer where the dynamics is
described by ideal ($\eta=\lambda=0$) and linear MHD. The linear
form of Eqs. (\ref{eq:masscontinuity1})-(\ref{eq:solenoid1}) can
be obtained by assuming a regular expansion of variables of the
form
\begin{equation}\label{eq:linearexpansion}
f=\epsilon f^{(1)}+\epsilon^{3/2} f^{(2)}\ldots,
\end{equation}
and collect only terms proportional to the small parameter
$\epsilon$. This leads to a system of linear equations for the
variables with superscript `1'. All variables can be eliminated
in favour of $u^{(1)}$ and $P^{(1)}$, leading to the system
\begin{equation}\label{eq:linearequation}
V\frac{\da P^{(1)}}{\da\theta}=F\frac{\da u^{(1)}}{\da
x},\phantom{X} V\frac{\da P^{(1)}}{\da x}=\rho_0 A\frac{\da
u^{(1)}}{\da\theta},
\end{equation}
where
\begin{equation}\label{eq:F}
F=\frac{\rho_0 A C}{V^4-V^2\left(v_A^2+c_S^2\right)+v_A^2
c_S^2\cos^{2}\alpha},
\end{equation}
\begin{equation}\label{eq:A}
A=V^2-v_A^2\cos^{2}\alpha,\nonumber
\end{equation}
\begin{equation}\label{eq:C}
C=\left(v_A^2+c_S^2\right)\left(V^2-c_T^2\cos^{2}\alpha\right).
\end{equation}
The quantities $A$ and $C$ vanish at the \alf and slow resonant
positions, respectively. As a result these two positions are regular
singular points for the system (\ref{eq:linearequation}). The
remaining variables can be expressed in terms of $u^{(1)}$ and
$P^{(1)}$ as,
\begin{equation}\label{eq:relationlinear1}
v_{\perp}^{(1)}=-\frac{V\sin\alpha}{\rho_0 A}P^{(1)}, \phantom{X}
v_{\parallel}^{(1)}=\frac{Vc_S^2\cos\alpha}{\rho_0 C}P^{(1)},
\end{equation}
\begin{equation}\label{eq:relationlinear2}
b_x^{(1)}=-\frac{B_0\cos\alpha}{V}u^{(1)},\phantom{X}
b_{\perp}^{(1)}=\frac{B_0\cos\alpha\sin\alpha}{\rho_0 A}P^{(1)},
\end{equation}
\begin{equation}\label{eq:relationlinear3}
\frac{\da b_{\parallel}^{(1)}}{\da\theta}=\frac{B_0\left(
V^2-c_S^2\cos^{2}\alpha\right)}{\rho_0 C}\frac{\da
P^{(1)}}{\da\theta} +\frac{u^{(1)}}{V}\frac{dB_0}{dx},
\end{equation}
\begin{equation}\label{eq:relationlinear4}
\frac{\da p^{(1)}}{\da\theta}=\frac{V^2 c_S^2}{C}
\frac{\da P^{(1)}}{\da\theta}-\frac{u^{(1)}B_0}{\mu_0 V}\frac{dB_0}{dx},
\end{equation}
\begin{equation}\label{eq:relationlinear5}
\frac{\da \rho^{(1)}}{\da\theta}=\frac{V^2}{C}
\frac{\da P^{(1)}}{\da\theta}+\frac{u^{(1)}}{V}\frac{d\rho_0}{dx}.
\end{equation}
Since Eq. (\ref{eq:linearequation}) has regular singular points, the
solutions can be obtained in terms of Fr\"{o}benius series with
respect to $x$ \citep[for details see, e.g.,][]{ruderman3, Ballai1998a} of the form
\begin{equation}\label{eq:frobenius1}
P^{(1)}=P_1^{(1)}(\theta)+P_2^{(1)}(\theta)x\ln|x|+P_3^{(1)}(\theta)+\ldots,
\end{equation}
\begin{equation}\label{eq:frobenius2}
u^{(1)}=u_1^{(1)}(\theta)\ln|x|+u_2^{(1)}(\theta)+u_3^{(1)}(\theta)x\ln|x|+\ldots.
\end{equation}
The coefficient functions depending on $\theta$ in the above
expansions are, generally, different for $x<0$ and $x>0$. The
particular form of these series solutions indicates that the
perturbation of the total pressure is regular at the ideal resonant
position. From Eqs.
(\ref{eq:relationlinear1})-(\ref{eq:relationlinear5}), we see that
the quantity $v_{\parallel}^{(1)}$ is also regular, while all other
quantities are singular. The quantities $u^{(1)}$, $b_x^{(1)}$,
$b_{\parallel}^{(1)}$, $p^{(1)}$ and $\rho^{(1)}$ behave as
$\ln|x|$, while $v_{\perp}^{(1)}$ and $b_{\perp}^{(1)}$ behave as
$x^{-1}$, so they are the most singular.

As the characteristic scale of dissipation is of the order of
$l_{inh}R^{-1/3}$ and we have assumed that $R\sim\epsilon^{3/2}$ we
obtain that the thickness of the dissipative layer is
$l_{inh}R^{-1/3}=\mathscr{O}(\epsilon^{1/2}l_{inh})$, implying the
introduction of a new stretched variable to replace the transversal
coordinate in the dissipative layer, which is defined as
$\xi=\epsilon^{-1/2}x$. Again, for brevity, Eqs.
(\ref{eq:masscontinuity1})-(\ref{eq:solenoid1}) are not rewritten as
they can be obtained by the substitution of
\begin{equation}\label{eq:subs}
\frac{\da}{\da x}=\epsilon^{-1/2}\frac{\da}{\da\xi},
\end{equation}
for all derivatives. The equilibrium quantities still depend on $x$,
not $\xi$ (their expression is valid in a wider region than the
characteristic thickness of the dissipative layer). All equilibrium quantities are
expanded around the ideal resonant position, $x=0$, as
\begin{equation}\label{eq:equilibsub}
f_0\approx f_{0_A}+\epsilon^{1/2} \xi\left(\frac{df_0}{dx}\right)_A,
\end{equation}
where $f_0$ is any equilibrium quantity and the subscript `A'
indicates that the equilibrium quantity is evaluated at the
Alfv\'{e}n resonant point.

We seek the solution to the set of equations obtained from Eqs.
(\ref{eq:masscontinuity1})-(\ref{eq:solenoid1}) by the substitution
of $x=\epsilon^{1/2}\xi$ into variables in the form of power series
of $\epsilon$. These equations contain powers of $\epsilon^{1/2}$,
so we use this quantity as an expansion parameter. To derive the
form of the inner expansions of different quantities we have to
analyze the outer solutions. First, since $v_{\parallel}$ and $P$
are regular at $x=0$ we can write their inner expansions in the
form of their outer expansions Eq. (\ref{eq:linearexpansion}). The
amplitudes of large variables in the dissipative layer are of the
order of $\epsilon^{1/2}$, so the inner expansion of the variables
$v_{\perp}$ and $b_{\perp}$ is
\begin{equation}\label{eq:innerexpansion}
g=\epsilon^{1/2}g^{(1)}+\epsilon g^{(2)}+\ldots.
\end{equation}
The quantities $u$, $b_x$, $b_{\parallel}$, $p$ and $\rho$ behave as
$\ln|x|$ in the vicinity of $x=0$, which suggests that they have
expansions with terms of the order of $\epsilon\ln\epsilon$ in the
dissipative layer. Strictly speaking, the inner expansions of all
variables have to contain terms proportional to
$\epsilon\ln\epsilon$ and $\epsilon^{3/2}\ln\epsilon$ \citep[see e.g.,][]{ruderman3}.
In the simplified version of matched
asymptotic expansions we utilize the fact that
$|\ln\epsilon|\ll\epsilon^{-\kappa}$ for any positive $\kappa$ and
$\epsilon\rightarrow +0$, and consider $\ln\epsilon$ as a quantity
of the order of unity \citep{Ballai1998a}. This enables us to write
the inner expansions for $u$, $b_x$, $b_{\parallel}$, $p$ and $\rho$
in the form of Eq. (\ref{eq:linearexpansion}).

We now substitute the expansion (\ref{eq:linearexpansion}) for $P$,
$u$, $b_x$, $b_{\parallel}$, $v_{\parallel}$, $p$ and $\rho$ and the
expansion given by Eq. (\ref{eq:innerexpansion}) for $v_{\perp}$ and
$b_{\perp}$ into the set of equations obtained from Eqs.
(\ref{eq:masscontinuity1})-(\ref{eq:solenoid1}) after substitution
of $x=\epsilon^{1/2}\xi$. The first order approximation (terms
proportional to $\epsilon$), yields a linear homogeneous system of
equations for the terms with superscript `1'. The important result
that follows from this set of equations is that
\begin{equation}\label{eq:pressuretheta}
P^{(1)}=P^{(1)}(\theta),
\end{equation}
that is $P^{(1)}$ does not change across the dissipative layer.
This result parallels the result found in linear theory
\citep{sakurai1, goossens1995b} and nonlinear theories of slow
resonance \citep[see e.g.,][]{ruderman3, Ballai1998a, Clack2008}.
Subsequently, all remaining variables can be expressed in terms of
$u^{(1)}$, $v_{\perp}^{(1)}$ and $P^{(1)}$ as
\begin{equation}
v_{\parallel}^{(1)}=\frac{c_{S}^2}{v_{A}^2}\frac{\cos\alpha}{\rho_{0}V}P^{(1)},
\end{equation}
\begin{equation}\label{eq:linearindisslayer}
b_\perp^{(1)}=-\frac{B_{0}V}{v_{A}^2\cos\alpha}v_\perp^{(1)},\mbox{ }b_x^{(1)}=-\frac{B_{0}\cos\alpha}{V}u^{(1)},
\end{equation}
\begin{equation}\label{eq:bparalinearrelation}
\frac{\da b_{\parallel}^{(1)}}{\da\theta}=\frac{B_{0}\left(v_{A}^2-c_{S}^2\right)}{\rho_{0}v_{A}^4}
\frac{dP^{(1)}}{d\theta}+\frac{u^{(1)}}{V}\left(\frac{dB_0}{dx}\right),
\end{equation}
\begin{equation}
\frac{\da p^{(1)}}{\da\theta}=\frac{c_{S}^2}{v_{A}^2}
\frac{dP^{(1)}}{d\theta}-\frac{u^{(1)}}{V}\frac{B_{0}}{\mu_0}\left(\frac{dB_0}{dx}\right),
\end{equation}
\begin{equation}
\frac{\da\rho^{(1)}}{\da\theta}=\frac{1}{v_{A}^2}\frac{dP^{(1)}}{d\theta}+\frac{u^{(1)}}{V}\left(\frac{d\rho_0}{dx}\right).
\end{equation}
All equilibrium quantities are calculated at $x=0$.
In addition, the
relation that connects the normal and perpendicular components of velocity is
\begin{equation}\label{eq:relationuvperp}
\frac{\da u^{(1)}}{\da\xi}-\sin\alpha\frac{\da v_{\perp}^{(1)}}{\da\theta}=0.
\end{equation}

In the second order approximation we only use the expressions
obtained from Eqs. (\ref{eq:momentumperp1}) and
(\ref{eq:inductperp1}). Employing Eqs.
(\ref{eq:pressuretheta})-(\ref{eq:relationuvperp}), we replace the
variables in the second order approximation which have superscript
`1'. The equations obtained in the second order are
\begin{multline}
\frac{\da
P^{(1)}}{\da\theta}\sin\alpha+\frac{B_{0}\cos\alpha}{\mu_0}\frac{\da
b_\perp^{(2)}}{\da\theta}
+V\rho_{0}\frac{\da v_\perp^{(2)}}{\da\theta}\\
=\frac{B_{0}V}{\mu_0 v_{A}^2}\left(\frac{dB_0}{dx}\right)\xi\frac{\da v_\perp^{(1)}}{\da\theta}
-V\left(\frac{d\rho_0}{dx}\right)\xi
\frac{\da v_\perp^{(1)}}{\da\theta}
-\eta\frac{\da^2v_\perp^{(1)}}{\da\xi^2},
\end{multline}
\begin{multline}
V\frac{\da b_\perp^{(2)}}{\da\theta}+B_{0}\cos\alpha\frac{\da
v_\perp^{(2)}}{\da\theta}
+\cos\alpha\left(\frac{dB_0}{dx}\right)\xi\frac{\da
v_\perp^{(1)}}{\da\theta}
=\lambda\frac{B_{0}V}{v_{A}^2\cos\alpha}\frac{\da^2v_\perp^{(1)}}{\da\xi^2}.\label{eq:rudermanpaper1161}
\end{multline}
Once the variables with superscript `2' have been eliminated from
the above two equations, the governing equation for resonant Alfv\'{e}n waves
inside the dissipative layer is derived as
\begin{align}\label{eq:clacky2}
    \triangle\xi\frac{\da v_\perp^{(1)}}{\da\theta} +
    \frac{V}{\rho_{0}}(\eta + \rho_{0}\lambda)\frac{\da^2 v_\perp^{(1)}}{\da\xi^2} =
    -\frac{V\sin\alpha}{\rho_{0}}\frac{dP^{(1)}}{d\theta},
\end{align}
where
\begin{equation}\label{eq:rudermanpaper1162}
\triangle=-\left(\frac{dv_A^2}{dx}\right)\cos^2\alpha.
\end{equation}

It is clear that Eq. (\ref{eq:clacky2}) does not contain nonlinear
terms despite considering the full MHD system of equations. This
result is in stark contrast with the results obtained for
nonlinear slow resonance where the governing equation was found to
be always nonlinear \citep[see e.g.,][]{ruderman3, Ballai1998a,
Clack2008}. The governing equation (\ref{eq:clacky2}) suggests that
resonant \alf waves can be described by the linear theory unless
their amplitudes inside the dissipative layer is of the order of
unity.

As the quadratic nonlinear terms cancel each other out, it is
natural to take into account cubic nonlinearity (the system of MHD
equations contain cubic nonlinear terms), where the nonlinearity
parameter is defined as
\begin{align}\label{eq:cubicnonlinearity}
    &\phi_{c} = \frac{g^2\da g/\da z}{\overline{\eta}\da^2 g/\da x^2}\simeq \epsilon^2 R.
\end{align}

Despite the higher order nonlinearity the governing equation is
similar to the equation derived for quadratic nonlinearity
(\ref{eq:clacky2}). These results require finding an explanation to
the linear behaviour of waves inside the dissipative layer. The
following section will be devoted to the study of nonlinear
corrections in the \alf dissipative layer.

\section{Nonlinear corrections in the \alf dissipative layer}

Since we have assumed that waves have small dimensionless amplitude
outside the dissipative layer, we will concentrate only on the
solutions \emph{inside} the dissipative layer.

In our analysis we use the assumptions and equations presented in
Section II, however, we will not impose any relation between $\epsilon$ and $R$. Equations
(\ref{eq:reycoefscales}) and (\ref{eq:transveralscale}) will be used
to define the scaled dissipative coefficients and stretching
transversal coordinate in the dissipative layer. For simplicity we
denote $\delta=R^{-1/3}$. This means that our scaled dissipative
coefficients and stretched transversal coordinate become
\begin{equation}\label{eq:delta1diff}
\overline{\eta}=\delta^3\eta,\mbox{
}\overline{\lambda}=\delta^3\lambda,\mbox{ }\xi=\delta^{-1}x.
\end{equation}
The first step to accomplish our task is to rewrite
Eqs. (\ref{eq:masscontinuity1})-(\ref{eq:solenoid1}) by substituting
\begin{align}\label{eq:subsxiandtheta}
&\frac{\da}{\da
x}=\delta^{-1}\frac{\da}{\da\xi},\mbox{ }\frac{\da}{\da z}=\frac{\da}{\da\theta},\mbox{ and }\frac{\da}{\da
t}=-V\frac{\da}{\da\theta}.
\end{align}
All equilibrium quantities (which are still dependent on $x$,
not $\xi$) will be approximated by the first non-vanishing term of
their Taylor expansion (see, Eq. (\ref{eq:equilibsub})).

The substitution of Eqs. (\ref{eq:equilibsub},
\ref{eq:delta1diff}-\ref{eq:subsxiandtheta}) will transform Eqs.
(\ref{eq:masscontinuity1})-(\ref{eq:solenoid1}) into
\begin{equation}\label{eq:massconservation1}
\rho_0\frac{\da u}{\da\xi}+\delta u\frac{d\rho_0}{dx}+\frac{\da(\rho
u)}{\da\xi}-\delta\frac{\da}{\da\theta}\left[\overline{\rho}\left(V-w\right)\right]=0,
\end{equation}
\begin{multline}\label{eq:momconservationnorm}
\frac{1}{\overline{\rho}}\left[\frac{\da
P}{\da\xi}-\frac{b_x}{\mu_0}\frac{\da
b_x}{\da\xi}-\frac{\delta}{\mu_0}\left(B_0\cos\alpha+b_z\right)\frac{\da
b_x}{\da\theta}\right]\\
=\delta\left(V-w\right)\frac{\da u}{\da\theta}-u\frac{\da
u}{\da\xi}+\delta^2\frac{\eta}{\rho_0}\frac{\da^2 u}{\da\xi^2},
\end{multline}
\begin{multline}\label{eq:momconservationperp}
\frac{1}{\overline{\rho}}\left[\delta\frac{\da
P}{\da\theta}\sin\alpha+\frac{b_x}{\mu_0}\frac{\da
b_{\perp}}{\da\xi}+\frac{\delta}{\mu_0}\left(B_0\cos\alpha+b_z\right)\frac{\da
b_{\perp}}{\da\theta}\right]\\
=-\delta\left(V-w\right)\frac{\da v_{\perp}}{\da\theta}+u\frac{\da
v_{\perp}}{\da\xi}-\delta^2\frac{\eta}{\rho_0}\frac{\da^2
v_{\perp}}{\da\xi^2},
\end{multline}
\begin{multline}\label{eq:momconservationpara}
\frac{1}{\overline{\rho}}\left[\delta\frac{\da
P}{\da\theta}\cos\alpha-\frac{b_x}{\mu_0}\frac{\da
b_{\parallel}}{\da\xi}-\frac{\delta}{\mu_0}\left(B_0\cos\alpha+b_z\right)\frac{\da
b_{\parallel}}{\da\theta}\right.\\
\left.-\frac{\delta}{\mu_0}\frac{dB_0}{dx}b_x\right]=
\delta\left(V-w\right)\frac{\da v_{\parallel}}{\da\theta}-u\frac{\da
v_{\parallel}}{\da\xi}+\delta^2\frac{4\eta}{\rho_0}\frac{\da^2
v_{\parallel}}{\da\xi^2},
\end{multline}
\begin{multline}\label{eq:inductnewnorm}
\delta\left(V-w\right)\frac{\da
b_x}{\da\theta}+\delta\left(B_0\cos\alpha+b_z\right)\frac{\da
u}{\da\theta}\\
+\delta^2\lambda\left(\frac{\da^2}{\da\xi^2}
+\delta^2\frac{\da^2}{\da\theta^2}\right)b_x=0,
\end{multline}
\begin{multline}\label{eq:inductnewperp}
\delta\left(V-w\right)\frac{\da b_{\perp}}{\da\theta}-u\frac{\da
b_{\perp}}{\da\xi}-b_{\perp}\left(\frac{\da
u}{\da\xi}+\delta\frac{\da
v_{\parallel}}{\da\theta}\cos\alpha\right)+b_x\frac{\da v_{\perp}}{\da\xi}\\
+\delta\left(B_0+b_{\parallel}\right)\frac{\da
v_{\perp}}{\da\theta}\cos\alpha
+\delta^2\lambda\left(\frac{\da^2}{\da\xi^2}+\delta^2\frac{\da^2}{\da\theta^2}\right)b_{\perp}=0,
\end{multline}
\begin{multline}\label{eq:inductnewpara}
\delta\left(V-w\right)\frac{\da
b_{\parallel}}{\da\theta}-u\left(\frac{\da
b_{\parallel}}{\da\xi}+\delta\frac{dB_0}{dx}\right)+b_{x}\frac{\da
v_{\parallel}}{\da\xi}
-\delta b_{\perp}\frac{\da v_{\parallel}}{\da\theta}\\
-\left(B_0+b_{\parallel}\right)\left(\frac{\da
u}{\da\xi}-\delta\frac{\da
v_{\perp}}{\da\theta}\sin\alpha\right)
+\delta^2\lambda\left(\frac{\da^2}{\da\xi^2}+\delta^2\frac{\da^2}{\da\theta^2}\right)b_{\parallel}=0,
\end{multline}
\begin{equation}\label{eq:adiabaticnew}
\left[\delta\left(V-w\right)\frac{\da}{\da\theta}-u\frac{\da}{\da\xi}\right]\left(\frac{\overline{p}}{\overline{\rho}^{\gamma}}\right)=0,
\end{equation}
\begin{equation}\label{eq:totalpressurenew}
P=p+\frac{B_0}{\mu_0}b_{\parallel}+\frac{1}{2\mu_0}\left(b_{x}^2+b_{\perp}^2+b_{\parallel}^2\right).
\end{equation}

The only condition we need to impose when deriving the nonlinear corrections
to resonant \alf waves in the dissipative layer is imported from the linear theory
which predicts that in the dissipative layer `large' variables have dimensionless
amplitude of the order of $\epsilon R^{1/3}$ (see Section III). We assume
that the dimensionless amplitudes of the linear approximation of `large' variables ($v_{\perp}$ and $b_{\perp}$)
in the dissipative layer are small, so that
\begin{equation}\label{eq:BASICASSUMPTION}
\epsilon\ll R^{-1/3}(=\delta).
\end{equation}
This condition ensures that the oscillation amplitude remains
small inside the dissipative layer. From a naive point of view
the linear theory is applicable as soon as the oscillation amplitude
is small. The example of slow resonant waves clearly shows that this
is not the case. The nonlinear effects become important in the slow
dissipative layer as soon as $\epsilon\sim R^{-2/3}$, i.e. as soon
as the oscillation amplitude in the dissipative layer, which is of
the order of $\epsilon R^{1/3}$, is of the order of $R^{-1/3}\ll1$.
For example, in the corona perturbations with dimensionless
amplitudes less than $10^{-4}$ can be considered by this theory.
From Eq. (\ref{eq:ratioofnonlindiss}) we would expect to see
quadratic nonlinearity appear for waves with dimensionless
amplitudes larger than $10^{-8}$ and from Eq.
(\ref{eq:cubicnonlinearity}) we would expect to see cubic
nonlinearity appear for waves with dimensionless amplitudes larger
than $10^{-6}$. If we take $\epsilon\approx R^{-1/3}$ we find that
inside the dissipative layer we have dimensionless amplitudes of the
order of unity. This causes a breakdown in our theory, and therefore
another approach would have to be adopted. At this time we do not
know of an analytical study which can carry out this task without
considering the full nonlinear MHD equations throughout the domain.

We now assume that all perturbations can be written as a regular
asymptotic expansion of the form
\begin{equation}\label{eq:regularexpansion}
\overline{f}=\overline{f}_0(x)+\epsilon \overline{f}_1(\xi,\theta)+\epsilon^2
\overline{f}_2(\xi,\theta)+\ldots,
\end{equation}
where $\overline{f}_0(x)$ represents the equilibrium value.
Substitution of expansion (\ref{eq:regularexpansion}) into the
system (\ref{eq:massconservation1})-(\ref{eq:totalpressurenew})
leads to a system of equations which contains the small parameter
$\delta$. This observation inspires us to look for the solution in
the form of expansions with respect to $\delta$. In order to cast
large and small variables in this description we are going to use
the following expansion for small variables ($u$, $b_x$,
$v_{\parallel}$, $b_{\parallel}$, $\rho$, $p$ and $P$)
\begin{equation}\label{eq:reynoldsexpansion}
\overline{g}_1=\overline{g}_{1}^{(1)}+\delta
\overline{g}_{1}^{(2)}+\ldots,
\end{equation}
while large variables
($v_{\perp}$ and $b_{\perp}$) will be expanded according to
\begin{equation}\label{eq:reynoldsexpansionlarge}
\overline{h}_1=\delta^{-1}\overline{h}_{1}^{(1)}+\overline{h}_{1}^{(2)}+\ldots.
\end{equation}
The bar notation is used here to distinguish between these
expansions and the expansions used in the previous section. From
this point on we drop the bar notation.

Substituting Eqs.
(\ref{eq:reynoldsexpansion})-(\ref{eq:reynoldsexpansionlarge}) into
the system (\ref{eq:massconservation1})-(\ref{eq:totalpressurenew}),
taking terms proportional to $\epsilon$ and then only retaining
terms with the lowest power of $\delta$, results in the set of
linear equations
\begin{equation}\label{eq:epsilonapprox1}
\rho_{0}\frac{\da
v_{\perp1}^{(1)}}{\da\theta}\sin\alpha-\rho_{0}\frac{\da
u_{1}^{(1)}}{\da\xi}=0,
\end{equation}
\begin{equation}\label{eq:epsilonapprox2}
\frac{\da P_1^{(1)}}{\da\xi}=0,
\end{equation}
\begin{equation}\label{eq:epsilonapprox3}
\rho_{0}V\frac{\da
v_{\perp1}^{(1)}}{\da\theta}+\frac{B_{0}\cos\alpha}{\mu_0}\frac{\da
b_{\perp1}^{(1)}}{\da\theta}=0,
\end{equation}
\begin{equation}\label{eq:epsilonapprox4}
\frac{\da P_1^{(1)}}{\da\theta}\cos\alpha-\rho_{0}V\frac{\da
v_{\parallel1}^{(1)}}{\da\theta}
-\frac{b_{x1}^{(1)}}{\mu_0}\left(\frac{dB_0}{dx}\right)
-\frac{B_{0}\cos\alpha}{\mu_0}\frac{\da
b_{\parallel1}^{(1)}}{\da\theta}=0,
\end{equation}
\begin{equation}\label{eq:epsilonapprox5}
V\frac{\da b_{x1}^{(1)}}{\da\theta}+B_{0}\cos\alpha\frac{\da
u_{1}^{(1)}}{\da\theta}=0,
\end{equation}
\begin{equation}\label{eq:epsilonapprox6}
V\frac{\da b_{\perp1}^{(1)}}{\da\theta}+B_{0}\cos\alpha\frac{\da
v_{\perp1}^{(1)}}{\da\theta}=0,
\end{equation}
\begin{equation}\label{eq:epsilonapprox7}
B_{0}\frac{\da u_1^{(1)}}{\da\xi}-B_{0}\sin\alpha\frac{\da
v_{\perp1}^{(1)}}{\da\theta}=0,
\end{equation}
\begin{equation}\label{eq:epsilonapprox8}
\rho_{0}V\frac{\da p_{1}^{(1)}}{\da\theta}-\rho_{0}c_{S}^2V\frac{\da\rho_1^{(1)}}{\da\theta}
+\rho_{0}u_1^{(1)}\left[c_{S}^2\left(\frac{d\rho_0}{dx}\right)-\left(\frac{dp_0}{dx}\right)\right]=0,
\end{equation}
\begin{equation}\label{eq:epsilonapprox9}
P_1^{(1)}-p_1^{(1)}-\frac{B_{0}}{\mu_0}b_{\parallel1}^{(1)}=0.
\end{equation}
In these equations all equilibrium quantities are calculated at $x=0$.

Using these equations we can express all dependent variables in
terms of $u_1^{(1)}$, $v_{\perp1}^{(1)}$ and $P_1^{(1)}$,
\begin{equation}
v_{\parallel1}^{(1)}=\frac{c_{S}^2}{v_{A}^2}\frac{\cos\alpha}{\rho_{0}V}P_1^{(1)},\label{eq:Fdone}
\end{equation}
\begin{equation}\label{eq:bperpvperpandubx}
b_{\perp1}^{(1)}=-\frac{B_{0}V}{v_{A}^2\cos\alpha}v_{\perp1}^{(1)},\mbox{ }
b_{x1}^{(1)}=-\frac{B_{0}\cos\alpha}{V}u_{1}^{(1)},
\end{equation}
\begin{equation}
\frac{\da b_{\parallel1}^{(1)}}{\da\theta}=\frac{B_{0}\left(v_{A}^2-c_{S}^2\right)}{\rho_{0}v_{A}^4}
\frac{dP_1^{(1)}}{d\theta}+\frac{u_1^{(1)}}{V}\left(\frac{dB_0}{dx}\right),\label{eq:Edone}
\end{equation}
\begin{equation}
\frac{\da p_{1}^{(1)}}{\da\theta}=\frac{c_{S}^2}{v_{A}^2}
\frac{dP_1^{(1)}}{d\theta}-\frac{u_1^{(1)}}{V}\frac{B_{0}}{\mu_0}\left(\frac{dB_0}{dx}\right),\label{eq:Gdone}
\end{equation}
\begin{equation}\label{eq:Ddone}
\frac{\da\rho_1^{(1)}}{\da\theta}=\frac{1}{v_{A}^2}\frac{dP_1^{(1)}}{d\theta}+\frac{u_1^{(1)}}{V}\left(\frac{d\rho_0}{dx}\right).
\end{equation}
It follows from Eq. (\ref{eq:epsilonapprox2}) that
\begin{equation}\label{eq:relP}
P_1^{(1)}=P_1^{(1)}(\theta).
\end{equation}
Finally, we obtain the relation between $u_1^{(1)}$ and
$v_{\perp1}^{(1)}$,
\begin{equation}\label{eq:reluvperp}
\frac{\da u_{1}^{(1)}}{\da\xi}=\frac{\da v_{\perp1}^{(1)}}{\da\theta}\sin\alpha.
\end{equation}

Note that Eqs. (\ref{eq:Fdone})-(\ref{eq:reluvperp}) are formally
identical to Eqs. (\ref{eq:pressuretheta})-(\ref{eq:relationuvperp})
for the linear approximation in Section III. This is not
surprising as both methods are designed to replicate linear theory
in the first order approximation.

\subsection{The second and third order nonlinear corrections}

Once the first order terms are known we can proceed to derive the
second and third order approximations with respect to $\epsilon$
(i.e. terms from the expansion of Eqs.
(\ref{eq:massconservation1})-(\ref{eq:totalpressurenew}) that are
proportional to $\epsilon^2$ and $\epsilon^3$, respectively). First,
we write out the second order approximations and substitute for all
first order terms (i.e. terms of the form $f_1^{(1)}$) using Eqs.
(\ref{eq:Fdone})-(\ref{eq:reluvperp}). Secondly, we find (by solving
the inhomogeneous system) the expansions of second order terms
(terms with \emph{subscript} `$2$'). Thirdly, we derive the second
order relations between all variables, similar to the ones obtained
in the first order approximation.

The equations representing the second order approximation with respect
to $\epsilon$ (with variables in the first order substituted) are
\begin{multline}\label{eq:epsilon2ndapprox1}
\rho_{0}\frac{\da
u_{2}}{\da\xi}+\delta\left[\xi\left(\frac{d\rho_0}{dx}\right)\frac{\da
u_2}{\da\xi}-V\frac{\da\rho_{2}}{\da\theta}+\rho_{0}\left(\frac{\da
v_{\parallel2}}{\da\theta}\cos\alpha\right.\right.\\
\left.\left.-\frac{\da v_{\perp2}}{\da\theta}\sin\alpha\right)
+\left(\frac{d\rho_0}{dx}\right)u_2\right]
=\frac{v_{\perp1}^{(1)}}{v_{A}^2}\frac{dP_{1}^{(1)}}{d\theta}\sin\alpha
+\mathscr{O}(\delta),
\end{multline}
\begin{equation}\label{eq:epsilon2ndapprox2}
\frac{\da P_2}{\da\xi}-\delta\left[\rho_{0}V\frac{\da
u_2}{\da\theta}+\frac{B_{0}\cos\alpha}{\mu_0}\frac{\da
b_{x2}}{\da\theta}\right]=\mathscr{O}(\delta),
\end{equation}
\begin{multline}\label{eq:epsilon2ndapprox3}
\frac{\da P_2}{\da\theta}\sin\alpha+\rho_{0}V\frac{\da
v_{\perp2}}{\da\theta} +\frac{B_{0}\cos\alpha}{\mu_0}\frac{\da
b_{\perp2}}{\da\theta}+
\delta\left\{\xi\left[V\left(\frac{d\rho_0}{dx}\right)\frac{\da
v_{\perp2}}{\da\theta}\right.\right.\\
\left.\left.+\frac{\cos\alpha}{\mu_0}\left(\frac{dB_0}{dx}\right)\frac{\da
b_{\perp2}}{\da\theta}\right]+\eta\frac{\da^2
v_{\perp2}}{\da\xi^2}\right\}=\mathscr{O}(\delta^{-1}),
\end{multline}
\begin{multline}\label{eq:epsilon2ndapprox4}
\frac{\da P_2}{\da\theta}\cos\alpha-\rho_{0}V\frac{\da
v_{\parallel2}}{\da\theta}
-\frac{b_{x2}}{\mu}\left(\frac{dB_0}{dx}\right)
-\frac{B_{0}\cos\alpha}{\mu_0}\frac{\da b_{\parallel2}}{\da\theta}\\
-\delta\left\{4\eta\frac{\da^2 v_{\parallel2}}{\da\xi^2}
+\xi\left[V\left(\frac{d\rho_0}{dx}\right)\frac{\da
v_{\parallel2}}{\da\theta}
+\frac{\cos\alpha}{\mu_0}\left(\frac{dB_0}{dx}\right)\frac{\da
b_{\parallel2}}{\da\theta}\right]\right\}\\
=\delta^{-1}\left[\frac{\cos\alpha\sin\alpha}{V}v_{\perp1}^{(1)}\frac{dP_1^{(1)}}{d\theta}\right]+\mathscr{O}(1),
\end{multline}
\begin{multline}\label{eq:epsilon2ndapprox5}
V\frac{\da b_{x2}}{\da\theta}+B_{0}\cos\alpha\frac{\da
u_{2}}{\da\theta}\\
+\delta\left[\lambda\frac{\da^2
b_{x2}}{\da\xi^2}
+\xi\cos\alpha\left(\frac{dB_0}{dx}\right)\frac{\da
u_{2}}{\da\theta}\right]=\mathscr{O}(1),
\end{multline}
\begin{multline}\label{eq:epsilon2ndapprox6}
V\frac{\da b_{\perp2}}{\da\theta}+B_{0}\cos\alpha\frac{\da
v_{\perp2}}{\da\theta}\\
+\delta\left[\lambda\frac{\da^2
b_{\perp2}}{\da\xi^2}
+\xi\cos\alpha\left(\frac{dB_0}{dx}\right)\frac{\da
v_{\perp2}}{\da\theta}\right]=\mathscr{O}(\delta^{-1}),
\end{multline}
\begin{multline}\label{eq:epsilon2ndapprox7}
B_{0}\frac{\da
u_2}{\da\xi}+\delta\left[\left(\frac{dB_0}{dx}\right)u_{2}-B_{0}\sin\alpha\frac{\da
v_{\perp2}}{\da\theta}-V\frac{\da
b_{\parallel2}}{\da\theta}\right.\\
\left.+\xi\left(\frac{dB_0}{dx}\right)\frac{\da u_2}{\da\xi}\right]=
\frac{B_{0}\sin\alpha}{\rho_{0}v_{A}^2}v_{\perp1}^{(1)}\frac{dP_1^{(1)}}{d\theta}+\mathscr{O}(\delta),
\end{multline}
\begin{equation}\label{eq:epsilon2ndapprox8}
\rho_{0}V\frac{\da
p_{2}}{\da\theta}-\rho_{0}c_{S}^2V\frac{\da\rho_2}{\da\theta}
+\rho_{0}u_2\left[c_{S}^2\left(\frac{d\rho_0}{dx}\right)-\left(\frac{dp_0}{dx}\right)\right]
=\mathscr{O}(\delta^{-1}),
\end{equation}
\begin{equation}\label{eq:epsilon2ndapprox9}
P_2-p_2-\frac{B_{0}}{\mu_0}b_{\parallel2}
+\delta\left[\frac{\xi}{\mu_0}\left(\frac{dB_0}{dx}\right)
b_{\parallel2}\right]
=\delta^{-2}\left[\frac{\rho_{0}}{2}{v_{\perp1}^{(1)}}^{2}\right]+\mathscr{O}(\delta^{-1}).
\end{equation}
It is clear that nonlinear terms appear from this order of
approximation and they are expressed in terms of variables obtained
in the first order.

The analysis of the system of Eqs.
(\ref{eq:epsilon2ndapprox1})-(\ref{eq:epsilon2ndapprox9}) reveals
that the expansions with respect to $\delta$ has to be written in
the form
\begin{equation}\label{eq:secondexpansion1}
g_2=\delta^{-1}g_2^{(1)}+g_2^{(2)}+\delta g_2^{(3)}+\ldots,
\end{equation}
for $u_2$, $b_{x2}$, $v_{\perp2}$, $b_{\perp2}$ and $P_2$ and
\begin{equation}\label{eq:secondexpansion2}
h_2=\delta^{-2}h_2^{(1)}+\delta^{-1}h_2^{(2)}+h_2^{(3)}+\ldots,
\end{equation}
for $v_{\parallel2}$, $b_{\parallel2}$, $p_2$ and $\rho_2$.

Here we need to make a note. It follows from Eqs.
(\ref{eq:secondexpansion1}) and (\ref{eq:secondexpansion2}) that
the ratio of $\rho_2$ to $\rho_1$ is of the order of
$\epsilon\delta^{-2}$, and the same is true for $v_{\parallel2}$,
$b_{\parallel2}$ and $p_2$. It seems to be inconsistent with the
regular perturbation method where it is assumed that the next
order approximation is always smaller than the previous one.
However, this problem is only apparent. To show this we need to
clarify the exact mathematical meaning of the statement ``in the
asymptotic expansion each subsequent term is much smaller than the
previous one". To do this we introduce the nine-dimensional vector
$\mathbf{U}=\left(u,v_{\parallel},v_{\perp},b_x,
b_{\parallel},b_{\perp},P,p,\rho\right)$ and consider it as an
element of a Banach space. The norm in this space can be
introduced in different ways. One possibility is
\begin{equation}\label{eq:normbanach}
||\mathbf{U}||=\int_{0}^{L}\mbox{}\rm{d}\theta\int_{-\infty}^{\infty}|\mathbf{U}|\mbox{ }\rm{d}\xi,
\end{equation}
where $L$ is the period. The asymptotic expansion in the dissipative layer, Eq. (\ref{eq:reynoldsexpansion}), can
be rewritten as
$\mathbf{U}=\mathbf{U}_0+\epsilon\mathbf{U}_1+\epsilon^2\mathbf{U}_2+\ldots$. Then the mathematical
formulation of the statement ``each subsequent term is much smaller than the previous one" is
$||\mathbf{U}_{n+1}||\ll||\mathbf{U}_n||$, $n=1,2,\ldots$. It is straightforward to verify that, in
accordance with Eqs. (\ref{eq:reynoldsexpansion}), (\ref{eq:reynoldsexpansionlarge}),
(\ref{eq:secondexpansion1}) and (\ref{eq:secondexpansion2}), $||\mathbf{U}_2||\ll||\mathbf{U}_1||$.

Once the expansions (\ref{eq:secondexpansion1}) and
(\ref{eq:secondexpansion2}) are substituted into Eqs.
(\ref{eq:epsilon2ndapprox1})-(\ref{eq:epsilon2ndapprox9}), we can
express the variables in this order of approximation as
\begin{equation}\label{eq:vpara2rel}
v_{\parallel2}^{(1)}=\frac{\cos\alpha}{2V}{v_{\perp1}^{(1)}}^2,
\quad
b_{\parallel2}^{(1)}=-\frac{B_{0}}{2v_{A}^2}{v_{\perp1}^{(1)}}^{2},
\end{equation}
\begin{equation}\label{eq:bvperp2bx2}
b_{\perp2}^{(1)}=v_{\perp2}^{(1)}=0,\quad
b_{x2}^{(1)}=-\frac{B_{0}\cos\alpha}{V}u_2^{(1)},
\end{equation}
\begin{equation}\label{eq:p2andrho2}
p_2^{(1)}=\rho_2^{(1)}=0.
\end{equation}
For the total pressure we obtain that
\begin{equation}\label{eq:P2}
\frac{\da P_2^{(1)}}{\da\xi}=0\Longrightarrow
P_2^{(1)}=P_2^{(1)}(\theta).
\end{equation}
In addition, we obtain that the equation which determines
$u_2^{(1)}$ is
\begin{equation}\label{eq:uvpara2}
\frac{\da u_2^{(1)}}{\da\xi}=-\frac{\cos^2\alpha}{V}v_{\perp1}^{(1)}\frac{\da v_{\perp1}^{(1)}}{\da\theta}.
\end{equation}

Since the large variables in this order of approximation are
$v_{\parallel2}$ and $b_{\parallel2}$, we can deduce that the linear
order of approximation of resonant \alf waves in the dissipative
layer excite \emph{magnetoacoustic} modes in the second order of
approximation. The excitation comes from the nonlinear term found in
the second order approximation of the pressure equation, this drives
the parallel components of the velocity and magnetic field
perturbations. Since we are focussed on the \alf resonance only,
these waves are not resonant. These waves act to cancel the very
small pressure and density perturbations created by the first order
approximation.

We now calculate the third order approximation with respect to
$\epsilon$. On analyzing the third order system of equations we
deduce that the large variables in this order of approximation are
Alfv\'{e}nic ($v_{\perp3}^{(1)}$, $b_{\perp3}^{(1)}$), so we only
need the perpendicular components of momentum and induction
equations given by Eqs. (\ref{eq:momconservationperp}) and
(\ref{eq:inductnewperp}), respectively.

First, since some of the first order approximation terms contribute
to the third order approximation in integral form we must introduce
a new notation
\begin{equation}\label{eq:U}
U_1^{(1)}=\int{u_1^{(1)}}d\theta.
\end{equation}
The third order approximation of the perpendicular component of
momentum is
\begin{multline}\label{eq:epsilon3rdapprox1}
\frac{\da P_3}{\da\theta}\sin\alpha+\rho_{0}V\frac{\da
v_{\perp3}}{\da\theta} +\frac{B_{0}\cos\alpha}{\mu_0}\frac{\da
b_{\perp3}}{\da\theta}+\\
\delta\left[\xi\left(V\left(\frac{d\rho_0}{dx}\right)\frac{\da
v_{\perp3}}{\da\theta}+\frac{\cos\alpha}{\mu_0}\left(\frac{dB_0}{dx}\right)\frac{\da
b_{\perp3}}{\da\theta}\right)+\eta\frac{\da^2
v_{\perp3}}{\da\xi^2}\right]\\
=\delta^{-2}\left\{\frac{\da
v_{\perp1}^{(1)}}{\da\xi}\left[
\frac{u_1^{(1)}P_1^{(1)}}{v_{A}^2}+\frac{u_1^{(1)}U_1^{(1)}}{V}
\left(\frac{d\rho_0}{dx}\right)\right]\right\}+\mathscr{O}(\delta^{-1}),
\end{multline}
while the perpendicular component of magnetic induction equation is
\begin{multline}\label{eq:epsilon3rdapprox2}
V\frac{\da b_{\perp3}}{\da\theta}+B_{0}\cos\alpha\frac{\da
v_{\perp3}}{\da\theta}+\delta\left[\lambda\frac{\da^2
b_{\perp3}}{\da\xi^2}\right.\\
\left.+\xi\cos\alpha\left(\frac{dB_0}{dx}\right)\frac{\da
v_{\perp3}}{\da\theta}\right]=\mathscr{O}(\delta^{-2}),
\end{multline}
Note that in obtaining the third order approximations we have
employed all the relations we have for variables in the first and
second order of approximation.

Equations (\ref{eq:epsilon3rdapprox1}) and
(\ref{eq:epsilon3rdapprox2}) clearly show that the nonlinear terms
on the right-hand sides do not cancel. This implies that the
expansion of $v_{\perp3}$ and $b_{\perp3}$ should be of the form
\begin{equation}\label{eq:vbperp3}
h_3=\delta^{-3}h_3^{(1)}+\delta^{-2}h_3^{(2)}+\delta^{-1}h_3^{(3)}+\ldots.
\end{equation}
We should state, for completeness, that if we derive the third order
approximation for all the Eqs. (\ref{eq:massconservation1})-(\ref{eq:totalpressurenew})
we obtain the expansions for $u_3$, $b_{x3}$, $v_{\parallel3}$, $b_{\parallel3}$,
$p_3$, $\rho_3$ and $P_3$ to be
\begin{equation}\label{eq:others3}
g_3=\delta^{-2}g_3^{(1)}+\delta^{-1}g_3^{(2)}+g_3^{(3)}+\ldots.
\end{equation}

The expansions calculated for all the variables
can now be collected together and we can write the
expansions for `large' and `small' variables in the dissipative layer
when studying resonant \alf waves.
Large variables ($v_{\perp}$ and $b_{\perp}$) have the expansion
\begin{equation}\label{eq:finalexpansionlarge}
h=\left(\frac{\epsilon}{\delta}\right)h_1^{(1)}+\epsilon\left(\frac{\epsilon}{\delta}\right)h_2^{(1)}
+\left(\frac{\epsilon}{\delta}\right)^3h_3^{(1)}+\ldots,
\end{equation}
and the expansion of small variables
($u$, $b_x$, $v_{\parallel}$, $b_{\parallel}$, $p$, $\rho$ and $P$)
is defined as
\begin{equation}\label{eq:finalexpansionsmall}
g=\epsilon g_1^{(1)}+\left(\frac{\epsilon}{\delta}\right)^2g_2^{(1)}
+\epsilon\left(\frac{\epsilon}{\delta}\right)^2g_3^{(1)}+\ldots.
\end{equation}

Since Eq. (\ref{eq:BASICASSUMPTION}) is the only condition enforced
in the dissipative layer, we can state that
\begin{equation}\label{eq:inequal}
1>\left(\frac{\epsilon}{\delta}\right)>\left(\frac{\epsilon}{\delta}\right)^2>\left(\frac{\epsilon}{\delta}\right)^3>\ldots.
\end{equation}

Therefore, since both Eqs. (\ref{eq:finalexpansionlarge}) and
(\ref{eq:finalexpansionsmall}) contain successive higher powers of
the parameter $\epsilon/\delta$ we can deduce that, considering Eq.
(\ref{eq:inequal}), higher orders of approximation of large and
small variables become increasingly insignificant in comparison to
the linear order of approximation, so resonant \alf waves in the
dissipative layer can be described accurately by linear theory if
condition (\ref{eq:BASICASSUMPTION}) is satisfied.

\section{Conclusions}

In the present paper we have investigated the nonlinear behaviour of
resonant \alf waves in the dissipative layer in one-dimensional
planar geometry in plasmas with anisotropic dissipative
coefficients, a situation applicable to solar coronal conditions.
The plasma motion outside the dissipative layer is described by the
set of linear, ideal MHD equations. The wave motion inside the
dissipative layer is governed by Eq. (\ref{eq:clacky2}). This
equation is linear, despite taking into consideration (quadratic and
cubic) nonlinearity. The Hall terms of the induction equation in the
perpendicular direction relative to the ambient magnetic field
cancel each other out.

The nonlinear corrections were calculated to explain why Eq.
(\ref{eq:clacky2}), describing the nonlinear behaviour of wave
dynamics, is always linear. We found that, in the second order of
approximation, magnetoacoustic modes are excited by the
perturbations of the linear order of approximation. These secondary
waves act to counteract the small pressure and density variations
created by the first order terms. In addition, these waves are not
resonant in the \alf dissipative layer. In the third order
approximation the perturbations become Alfv\'{e}nic, however, these
perturbations are much smaller than those in the linear order of
approximation. Equations (\ref{eq:finalexpansionlarge}) and
(\ref{eq:finalexpansionsmall}) describe the expansion of large and
small variables, respectively, and demonstrate that all higher order
approximations of both large and small variables at the Alfv\'{e}n
resonance are smaller than the linear order approximation, provided
condition (\ref{eq:BASICASSUMPTION}) is satisfied. This condition
ensures that the oscillation amplitude remains small inside the
dissipative layer. From a naive point of view the linear theory is
applicable as soon as the oscillation amplitude is small. The
example of slow resonant waves clearly shows that this is not the
case. The nonlinear effects become important in the slow dissipative
layer as soon as $\epsilon\sim R^{-2/3}$, i.e. as soon as the
oscillation amplitude in the dissipative layer, which is of the
order of $\epsilon R^{1/3}$, is of the order of $R^{-1/3}\ll1$. We
also found that any dispersive effect due to the consideration of
ions' inertial length (Hall effect) is absent from the governing
equation.

This calculation of nonlinear corrections to resonant \alf waves in
dissipative layers allows us to apply the already well-known linear
theory for studying resonant \alf waves in the solar corona with
great accuracy, where the governing equation, jump conditions and
the absorption of wave energy are already derived \citep[see e.g.,][]{sakurai1, goossens1995b, erdelyi1998}.

It is interesting to note that this work can be transferred to
isotropic plasma rather easily. Shear viscosity, supplied by
Braginskii's viscosity tensor (see Appendix A), acts exactly as
isotropic viscosity. Therefore, replacing $\eta$ by
$\rho_{0_a}\nu$ in Eq. (\ref{eq:clacky2}) provides the required
governing equation for resonant \alf waves in isotropic plasmas.
Moreover, the work on the nonlinear corrections presented in this
paper is also unaltered by anisotropy. This implies that we can
consider resonant \alf waves in dissipative layers throughout the
solar atmosphere and still use linear theory if condition
(\ref{eq:BASICASSUMPTION}) is satisfied.

\begin{acknowledgements}

C.T.M.C. would like to thank STFC (Science and Technology Facilities
Council) for the financial support provided. I.B. was financially
supported by NFS Hungary (OTKA, K67746) and The National University
Research Council Romania (CNCSIS-PN-II/531/2007)
\end{acknowledgements}

\begin{appendix}
\section{Braginskii's viscosity tensor and derivation of largest terms}

In this Appendix, we shall derive the largest terms of Braginskii's
viscosity tensor inside the \alf dissipative layer to be used to
study the nonlinearity effects. Braginskii's viscosity tensor
comprises of five terms. Its divergence can be written as
\citep{braginskii}
\begin{equation}\label{eq:braginskiistensor}\tag{A1}
\nabla\cdot\mathbf{S}=\overline{\eta}_0\nabla\cdot\mathbf{S}_0+
\overline{\eta}_1\nabla\cdot\mathbf{S}_1+\overline{\eta}_2\nabla\cdot\mathbf{S}_2
-\overline{\eta}_3\nabla\cdot\mathbf{S}_3-\overline{\eta}_4\nabla\cdot\mathbf{S}_4,
\end{equation}
Note that the terms proportional to $\overline{\eta}_0$,
$\overline{\eta}_1$ and $\overline{\eta}_2$ in Eq.
(\ref{eq:braginskiistensor}) describe viscous dissipation, while
terms proportional to $\overline{\eta}_3$ and $\overline{\eta}_4$
are non-dissipative and describe the wave dispersion related to the finite
ion gyroradius. They will be ignored in what follows. For simplicity in the body of the paper,
we have taken $\eta=\eta_1$.

The quantities $\mathbf{S}_0$, $\mathbf{S}_1$ and $\mathbf{S}_2$ are
given by
\begin{equation}\label{eq:S0}\tag{A2}
\mathbf{S}_0=\left(\mathbf{b}'\otimes\mathbf{b}'-\frac{1}{3}I\right)\left[3\mathbf{b}'
\cdot\nabla(\mathbf{b}'\cdot\mathbf{v})-\nabla\cdot\mathbf{v}\right],
\end{equation}
\begin{multline}\label{eq:S1}\tag{A3}
\mathbf{S}_1=\nabla\otimes\mathbf{v}+\left(\nabla\otimes\mathbf{v}\right)^T-\mathbf{b}'\otimes\mathbf{W}
-\mathbf{W}\otimes\mathbf{b}'\\
+\left(\mathbf{b}'\otimes\mathbf{b}'-I\right)\nabla\cdot\mathbf{v}
+\left(\mathbf{b}'\otimes\mathbf{b}'+I\right)\mathbf{b}'\cdot\nabla\left(\mathbf{b}'\cdot\mathbf{v}\right),
\end{multline}
\begin{equation}\label{eq:S2}\tag{A4}
\mathbf{S}_2=\mathbf{b}'\otimes\mathbf{W}+\mathbf{W}\otimes\mathbf{b}'
-4\left(\mathbf{b}'\otimes\mathbf{b}'\right)\mathbf{b}'\cdot\nabla\left(\mathbf{b}'\cdot\mathbf{v}\right),
\end{equation}
\begin{equation}\label{eq:W}\tag{A5}
\mathbf{W}=\nabla\left(\mathbf{b}'\cdot\mathbf{v}\right)
+\left(\mathbf{b}'\cdot\nabla\right)\mathbf{v}.
\end{equation}
Here $\mathbf{v}=(u,v,w)$ is the velocity,
$\mathbf{b}'=\mathbf{B}_0/B_0$, $\mathbf{I}$ is the unit tensor and $\otimes$
indicates the dyadic product of two vectors. The superscript `T'
denotes a transposed tensor.

The first viscosity coefficient, $\overline{\eta}_0$,
(\emph{compressional viscosity}) has the following approximate
expression \citep[see e.g.,][]{ruderman2000b}
\begin{equation}\label{eq:approxeta0}\tag{A6}
\overline{\eta}_0=\frac{\rho_0k_BT_0\tau_i}{m_p},
\end{equation}
where $\rho_0$ and $T_0$ are the equilibrium density and pressure,
$m_p$ is the proton mass, $k_B$ the Boltzmann constant and $\tau_i$
the ion collision time. The other viscosity coefficients depend on
the quantity $\omega_i\tau_i$, where $\omega_i$ it the ion
gyrofrequency. When $\omega_i\tau_i\gg1$ these coefficients are
given by the approximate expressions
\begin{equation}\label{eq:coeffapproximates1}\tag{A7}
\overline{\eta}_1=\frac{\overline{\eta}_0}{4\left(\omega_i\tau_i\right)^{2}},
\mbox{ }\overline{\eta}_2=4\overline{\eta}_1.
\end{equation}

The viscosity described by the sum of the second and third terms
in Eq. (\ref{eq:braginskiistensor}) is \emph{the shear viscosity}.
For typical coronal conditions $\omega_i\tau_i$ is of the order of
$10^5-10^6$, so according to Eq. (\ref{eq:coeffapproximates1}) the
term proportional to $\overline{\eta}_0$ in Eq.
(\ref{eq:braginskiistensor}) is much larger than the second and
third terms. However, it has been long understood that the
compressional viscosity does not remove the \alf singularity
\citep[see e.g.,][]{erdelyi1995, mocanu2008} while shear viscosity
does.

First, we shall calculate the components of the compressional
viscosity. We will use the notation of parallel and perpendicular
components as defined in the paper. It is straightforward to obtain
that
\begin{equation}\label{compressx}
\overline{\eta}_0\left(\nabla\cdot\mathbf{S}_0\right)_x=0,\tag{A8}\\
\end{equation}
\begin{equation}\label{compressperp}
\overline{\eta}_0\left(\nabla\cdot\mathbf{S}_0\right)_{\perp}=0,\tag{A9}\\
\end{equation}
\begin{equation}\label{compresspara}
\overline{\eta}_0\left(\nabla\cdot\mathbf{S}_0\right)_{\parallel}=
\overline{\eta}_0\cos\alpha\left(2\frac{\da^2 v_{\parallel}}{\da
z^2}\cos\alpha-\frac{\da^2 u}{\da x\da z}+\frac{\da^2 v_{\perp}}{\da
z^2}\sin\alpha\right).\tag{A10}
\end{equation}
The shear viscosity, as stated above, is the sum of the second and
third terms of Eq. (\ref{eq:braginskiistensor}). To evaluate these terms we use
the approximate expression for $\overline{\eta}_1$ and
$\overline{\eta}_2$ given by Eq. (\ref{eq:coeffapproximates1}).
As a result we obtain
\begin{multline}\label{eq:shearx}\tag{A11}
\overline{\eta}_1\left[\left(\nabla\cdot\mathbf{S}_1\right)_x
+4\left(\nabla\cdot\mathbf{S}_2\right)_x\right]=
\overline{\eta}_1\left[\frac{\da^2 u}{\da x^2}\right.\\
\left.+(1+3\cos^2\alpha)\frac{\da^2 u}{\da z^2} +4\frac{\da^2
v_{\parallel}}{\da x \da z}\cos\alpha\right],
\end{multline}
\begin{multline}\label{eq:shearperp}\tag{A12}
\overline{\eta}_1\left[\left(\nabla\cdot\mathbf{S}_1\right)_{\perp}
+4\left(\nabla\cdot\mathbf{S}_2\right)_{\perp}\right]=\\
\overline{\eta}_1\left[ \frac{\da^2 v_{\perp}}{\da x^2}\right.
\left.+\left(4-3\sin^2\alpha-16\sin^6\alpha\right)\frac{\da^2
v_{\perp}}{\da z^2}\right.\\
\left.+4\sin\alpha\cos\alpha\left(4\sin^4\alpha-1\right)\frac{\da^2
v_{\parallel}}{\da z^2}\right],
\end{multline}
\begin{multline}\label{eq:shearpara}\tag{A13}
\overline{\eta}_1\left[\left(\nabla\cdot\mathbf{S}_1\right)_{\parallel}
+4\left(\nabla\cdot\mathbf{S}_2\right)_{\parallel}\right]=\\
4\overline{\eta}_1\left\{\frac{\da^2 v_{\parallel}}{\da
x^2}+\left[1+\cos^2\alpha\left(4\sin^4\alpha-1\right)\right]\frac{\da^2
v_{\parallel}}{\da z^2}\right.\\
\left.+\frac{\da^2 u}{\da x\da
z}\cos\alpha-\left(4\sin^4\alpha+1\right)\frac{\da^2 v_{\perp}}{\da
z^2}\cos\alpha\sin\alpha\right\}.
\end{multline}

Equations (\ref{compressx})-(\ref{eq:shearpara}) are complicated,
but we can simplify them further by taking the largest term only in
each equation. For the viscosity in the parallel direction it would,
at first, seem obvious that the largest term will be proportional to
$\overline{\eta}_0$ rather than $\overline{\eta}_1$. However, some
of the variables proportional to $\overline{\eta}_1$ have
derivatives with respect to $x$ which produce enormous gradients in
the dissipative layer when there is a transversal inhomogeneity,
hence some of the terms proportional to $\overline{\eta}_1$ are of
the same order as or larger than the terms proportional to
$\overline{\eta}_0$. It is also important to note that in the first
order approximation the second and third terms on the right-hand
side of Eq. (\ref{compresspara}) cancel (see Eq.
\ref{eq:relationuvperp}). For the normal and perpendicular
components of viscosity, the treatment is slightly simpler. The
compressional viscosity is zero, and as derivatives with respect to
$z$ are much small that derivatives with respect to $x$, we can
select the largest term proportional to $\overline{\eta}_1$ by
observation. Therefore, the viscosity tensor can be approximated by
\begin{equation}\label{eq:lrgeviscx}\tag{A14}
\left(\nabla\cdot\mathbf{S}\right)_x\approx\overline{\eta}_1\frac{\da^2
u}{\da x^2},
\end{equation}
\begin{equation}\label{eq:lrgeviscperp}\tag{A15}
\left(\nabla\cdot\mathbf{S}\right)_{\perp}\approx\overline{\eta}_1\frac{\da^2
v_{\perp}}{\da x^2},
\end{equation}
\begin{equation}\label{eq:lrgeviscpara}\tag{A16}
\left(\nabla\cdot\mathbf{S}\right)_{\parallel}\approx4\overline{\eta}_1\frac{\da^2
v_{\parallel}}{\da x^2}.
\end{equation}

Equations (\ref{eq:lrgeviscx})-(\ref{eq:lrgeviscpara}) give an
appropriate approximation to Braginskii's viscosity tensor when
studying nonlinear resonant \alf waves in dissipative layers. It is interesting to note that
the terms in Eqs. (\ref{eq:lrgeviscx})-(\ref{eq:lrgeviscpara}) are
identical to the largest terms when considering isotropic viscosity.
Obviously, compressional viscosity cannot remove the \alf singularity
since Eq. (\ref{compressperp}) is identically zero.
\end{appendix}

\begin{appendix}
\section{The derivation of the Hall term in the induction equation
for Alfv\'{e}n resonant waves}

In this Appendix we will derive the components of the Hall term in
the induction equation and show that neglecting the Hall effect at
the \alf resonance is a good approximation for typical conditions
throughout the solar atmosphere. The main reasons qualitatively are
as follows. When we are in the lower solar atmosphere (e.g. solar
photosphere) the Hall conduction is much smaller than the direct
conduction since the product of the electron gyrofrequency,
$\omega_e$, and collision time, $\tau_e$, is less than unity
\citep[see e.g.,][]{priest1}. For the upper atmosphere (e.g.
chromosphere, corona), where the product $\omega_e\tau_e$ is greater
than unity, the Hall conduction has to be considered. However, when
the Hall terms are derived, the largest terms in the perpendicular
direction relative to the ambient magnetic field cancel leaving only
higher order approximation terms which are far smaller than the
direct conduction. As the dominant dynamics of resonant \alf waves
in dissipative layer resides in the components of velocity and
magnetic field perturbation in the perpendicular direction relative
to the background magnetic field we can neglect the Hall conduction
completely from the analysis without affecting the governing
equation.

In order to estimate the relative importance of the Hall term and
resistive term in the dissipative layer we follow the
sophisticated analysis which was presented by \citet{ruderman3} and \citet{Clack2008}.
We do not write down all the steps of the analysis, but rather
give the salient points specific to the Hall effect at the Alfv\'{e}n resonance.

Equations (\ref{eq:linearexpansion}) and (\ref{eq:innerexpansion})
provide the following estimations in the dissipative layer:
\begin{align}\label{eq:estimates}
&u=\mathscr{O}(\epsilon),\phantom{x}v_{\perp}=\mathscr{O}(\epsilon^{1/2}),\phantom{x}
v_{\parallel}=\mathscr{O}(\epsilon),\nonumber\\
&b_x=\mathscr{O}(\epsilon),\phantom{x}b_{\perp}=\mathscr{O}(\epsilon^{1/2})
\phantom{x}b_{\parallel}=\mathscr{O}(\epsilon),\tag{B1}
\end{align}
where $\epsilon$ still denotes the dimensionless amplitude of
oscillations far away from the dissipative layer.
The thickness of the dissipative layer divided by the characteristic
scale of inhomogeneity is
$\delta_c/l_{inh}=\mathscr{O}(\epsilon^{1/2})$. This gives rise to
the estimations
\begin{equation}\label{eq:inhomoq}
l_{inh}\frac{\da h}{\da
x}=\mathscr{O}(\epsilon^{-1/2}h),\quad
l_{inh}\frac{\da h}{\da z}=\mathscr{O}(h),\quad
l_{inh}^{2}\frac{\da^2 h}{\da z^2}=\mathscr{O}(h),\tag{B2}
\end{equation}
where $h$ denotes any of the quantities $u$, $b_x$, $b_{\parallel}$, $b_{\perp}$ or $v_{\perp}$.
Since the first term in the expansion of $v_\parallel$ is independent of $x$, it follows that
\begin{equation}\label{eq:inhomobperp}
l_{inh}\frac{\da v_{\parallel}}{\da
x}=\mathscr{O}(v_{\parallel}),\quad
l_{inh}\frac{\da v_{\parallel}}{\da z}=\mathscr{O}(v_{\parallel}),\quad
l_{inh}^{2}\frac{\da^2 v_{\parallel}}{\da
x^2}=\mathscr{O}(\epsilon^{-1/2}v_{\parallel}).\tag{B3}
\end{equation}

We now need to calculate the components of the vectors of the
resistive term and the Hall term from the induction equation
normal to the magnetic surfaces (the
$x$-direction) and in the magnetic surfaces parallel and
perpendicular to the equilibrium magnetic field lines. We use Eqs.
(\ref{eq:inhomoq}) and (\ref{eq:inhomobperp}) in order to estimate
all the terms and we only retain the largest. As a result we have
\begin{align}
&\overline{\lambda}\nabla^2 B_x=\overline{\lambda}\frac{\da^2 b_x}{\da x^2}+\ldots,\tag{B4}\\
&\overline{\lambda}\nabla^2 B_{\perp}=\overline{\lambda}\frac{\da^2 b_{\perp}}{\da x^2}+\ldots,\tag{B5}\\
&\overline{\lambda}\nabla^2 B_{\parallel}=\overline{\lambda}\frac{\da^2 b_{\parallel}}{\da x^2}+\ldots,\tag{B6}\\
&H_x=\frac{B_0\cos^2\alpha}{\mu_0 en_e}\frac{\da^2 b_{\perp}}{\da z^2}+\ldots,\tag{B7}\\
&H_{\perp}=\frac{B_0}{\mu_0 en_e}\left(\frac{1}{B_0}\frac{dB_0}{dx}\frac{\da b_{x}}{\da x}+\cos\alpha\frac{\da^2 b_{\parallel}}{\da z\da x}\right)+\ldots,\tag{B8}\label{eq:b14}\\
&H_{\parallel}=-\frac{B_0\cos\alpha}{\mu_0 en_e}\frac{\da^2 b_{\perp}}{\da z\da x}+\ldots,\tag{B9}
\end{align}
where the dots indicate terms much smaller than those shown explicitly.
With the aid of Eqs. (\ref{eq:estimates}), (\ref{eq:inhomoq}) and
(\ref{eq:inhomobperp}) we obtain the ratios
\begin{align}
&\frac{H_x}{\overline{\lambda}\nabla^2B_x}\sim\epsilon^{1/2}\omega_e\tau_e,\tag{B10}\label{eq:xrat}\\
&\frac{H_{\perp}}{\overline{\lambda}\nabla^2B_{\perp}}\sim\epsilon\omega_e\tau_e,\tag{B11}\label{eq:perprat}\\
&\frac{H_{\parallel}}{\overline{\lambda}\nabla^2B_{\parallel}}\sim\omega_e\tau_e.\tag{B12}\label{eq:pararat}
\end{align}
For the Hall conduction to be significant in the direction of the dominant dynamics of resonant
\alf waves (i.e. in the perpendicular direction) we must have $\epsilon\omega_e\tau_e\gtrsim1$.
This is plausible for the solar upper atmosphere. If this condition holds, then we must consider the
Hall term in the induction equation. However, if Eq. (\ref{eq:b14}) is expanded using Eq. (\ref{eq:linearexpansion})
from Section III we obtain
\begin{equation}\tag{B13}\label{eq:hall1}
\epsilon^{3/2}\lambda\left\{\frac{1}{B_{0}}\left(\frac{dB_0}{dx}\right)\frac{\da b_x^{(1)}}{\da\xi}
+\cos\alpha\frac{\da^2 b_{\parallel}^{(1)}}{\da\theta\da\xi}\right\}+\mathscr{O}(\epsilon^{2}).
\end{equation}
It should be noted that in deriving Eq. (\ref{eq:hall1}) we have
used the assumption that $\epsilon\omega_e\tau_e=\mathscr{O}(1)$.
The terms inside the braces are of the same order as the direct
conduction. Hence, they would be expected to appear in the governing
equation. When substituting for $b_x^{(1)}$ and
$b_{\parallel}^{(1)}$ using Eqs. (\ref{eq:linearindisslayer}) and
(\ref{eq:bparalinearrelation}), respectively, it is found that the
terms inside the brackets cancel
\begin{equation}\tag{B14}\label{cancelhall}
\frac{\cos\alpha}{V}\left(\frac{dB_0}{dx}\right)\frac{\da u^{(1)}}{\da\xi}
-\frac{\cos\alpha}{V}\left(\frac{dB_0}{dx}\right)\frac{\da u^{(1)}}{\da\xi}=0.
\end{equation}

Equation (\ref{cancelhall}) shows that the Hall term in the perpendicular component
of induction is always smaller than the direct conduction in the solar atmosphere.
The normal and parallel components of the Hall conduction are, in fact, larger than the
perpendicular component. Nevertheless they play no role in derivation of the governing
equation of resonant \alf waves in dissipative layer. The parallel component is the largest
of the three components and this is to be expected as the Hall effect is strongest at right
angles to the dominant wave motion. This is in complete agreement with the study on resonant slow waves by
\citet{Clack2008} which found the largest Hall effect was in the perpendicular component of
Hall conduction, which is at right angles to the dominant dynamics of resonant slow waves.

In summary, it is a good approximation to neglect the Hall term in the induction equation
when studying resonant \alf waves in dissipative layer. This approximation holds throughout
the entire solar atmosphere.
\end{appendix}

\bibliographystyle{aa} % style aa.bst
\bibliography{aa11106_ccibmr}

\end{document}